\documentclass[aps,preprint,showpacs,preprintnumbers,showkeys]{revtex4-1}
\usepackage{dcolumn}

\usepackage[usenames,dvipsnames]{color}

\usepackage{graphicx}
\usepackage{epstopdf}

\usepackage{amssymb}
\usepackage{amsmath}
\usepackage{bm}
\usepackage{epsfig}
\usepackage{ulem}

\newcommand {\be}{\begin{equation}}
\newcommand {\ee}{\end{equation}}
\newcommand {\bea}{\begin{eqnarray}}
\newcommand {\eea}{\end{eqnarray}}
\newcommand {\FIG}[1]{Fig. \ref{#1}}
\newcommand {\FIGS}[1]{Figs. \ref{#1}}

\newcommand {\PRE}[1]{{Phys. Rev. E} {\bf {#1}}}

\newcommand {\PRL}[1]{{Phys. Rev. Lett.} {\bf {#1}}}
\newcommand {\EPL}[1]{{Europhys. Lett.} {\bf {#1}}}

\newcommand {\SCI}[1]{{Science} {\bf {#1}}}

\begin{document}

\title{Agglomerative percolation  on the Bethe lattice and the triangular cactus}
\author{Huiseung Chae}
\author{Soon-Hyung Yook}\email{syook@khu.ac.kr}
\author{Yup Kim} \email[Corresponding author:]{ykim@khu.ac.kr  }
\affiliation{Department of Physics and Research Institute for Basic
Sciences, Kyung Hee University, Seoul 130-701, Korea}
\date{\today}

\begin{abstract}
We study the agglomerative percolation (AP) models  on the Bethe lattice and the triangular cactus to
establish the exact mean-field theory for AP. Using the self-consistent simulation method, based on the
exact self-consistent equation, we directly measure the order parameter $P_{\infty}$ and average
cluster size $S$. From the measured $P_{\infty}$ and $S$ we obtain the critical exponents $\beta_k$
and $\gamma_k$ for $k=2$ and $3$. Here $\beta_k$ and $\gamma_k$ are the critical exponents for $P_\infty$ and $S$ 
when the growth of clusters spontaneously breaks the $Z_k$ symmetry of the $k$-partite graph  \cite{Z2SYM}.
The obtained values are $\beta_2=1.79(3)$, $\gamma_2=0.88(1)$,
$\beta_3=1.35(5)$, and $\gamma_3=0.94(2)$. By comparing these values of exponents with those for ordinary
percolation ($\beta_{\infty}=1$ and $\gamma_{\infty}=1$) we also find the inequalities between the exponents, as
$\beta_\infty<\beta_3<\beta_2$ and $\gamma_\infty>\gamma_3>\gamma_2$.
These results quantitatively verify  the conjecture that the AP model belongs to a new universality class if $Z_k$ symmetry is broken spontaneously,
and the new universality class depends on $k$ [Lau {\it et al.}, Phys. Rev. E {\bf 86}, 011118 (2012)] .
\end{abstract}
\maketitle

\section{Introduction}

Percolation transition describes the emergence of large-scale connectivity \cite{Stauffer_book}.
It has been extensively studied in various branches of science
due to its wide range of applications
to many phenomena such as sol-gel transition and polymerization,
resistor networks, and epidemic spreading \cite{Stauffer_book}.
The first theoretical model for the percolation
is the random or ordinary percolation in which
a vacant site or a vacant bond of the background lattice is randomly chosen to be occupied. 
The percolation transition in the random percolation is normally known to be
continuous \cite{Stauffer_book}. The percolation has been extensively studied during last 3 or 4 decades to be considered as a mature branch of sciences.

However anomalous physical properties of exotic models on the percolation recently triggered some new studies. One kind of studies \cite{Ach} was on the explosive percolation, which was first known to show supposedly discontinuous transition on the complete graph (CG) \cite{Ach,DSouza}. But subsequent studies on the explosive percolation 
have shown that the transition of the explosive percolation on CG or the mean-field transition is continuous \cite{daCosta10,Lee11,Rio,Chae0}.

Another kind of studies was on the agglomerative percolation (AP) \cite{1DRING, 2Ds, ERN, Z2SYM, CRIT}. In AP one cluster is randomly selected 
instead of a bond or a site. Then the selected cluster merges all the nearest neighboring clusters to form a new cluster.
The phase transition in AP is shown to be continuous, but belongs to a new universality class different from the class
of the random percolation if the base structure of AP is bipartite \cite{Z2SYM}.
On the bipartite structure like a two-dimensional square lattice the merging process spontaneously breaks the $Z_2$
symmetry at the transition threshold, 
which is the origin of the new universality class \cite{Z2SYM}.
In contrast, the universality of the transition of AP on the triangular lattice, which is not bipartite, is the same as that of the random percolation \cite{Z2SYM}.
Using analytical methods and numerical simulations, APs on the one-dimensional ring \cite{1DRING},
the two-dimensional square lattice and triangular lattice \cite{2Ds}, critical tree \cite{CRIT}, and complex network \cite{ERN} were studied. Through these studies AP on bipartite graphs is shown to belong to a new universality class different from that of the random percolation. 

To understand and establish a new universality class of the critical phenomena clearly
and precisely the exact mean-field theory for the new model must be first understood.  However the mean-field theory (MFT) for AP on bipartite networks was not clearly understood yet.
To get MFT of AP, the analytic theory based on the generating function of the Erd\"{o}s-R\'{e}nyi (ER) random network 
was attempted \cite{ERN}. However this analytic approach predicted the critical
exponent $\gamma$ as $\gamma=1/2$, but from the numerical simulation on ER graph 
$\gamma=0.88(10)$ is obtained, which is significantly larger than $\gamma=1/2$.
This suggests a possibility that the analytic approach based on the generating function \cite{ERN} is still far from completion.
Furthermore ER graph is not exactly bipartite.
The numerical simulation study on the exact bipartite
random graph earned only the critical exponent $\nu$ and the fractal dimension $D$ of the giant cluster
as $\nu=4.7(2)$ and $D=0.567(6)$, which are close to those for ER network but differ by more than one standard deviation \cite{Z2SYM}.
Therefore, at the present stage, MFT for AP on the bipartite graphs are far from completion.
  
Recently the complete graph is widely used as a testbed  for MFT \cite{Ach,daCosta10,Lee11,Rio}.
However the complete graph is not bipartite and one growth step of AP on the complete graph makes the entire graph
a new single cluster.
In contrast  the Bethe lattice (infinite homogeneous Cayley tree)
is the exact bipartite graph on which AP can be  well defined.
Moreover, the Bethe lattice  is physically a very
important substrate or medium on which
MFTs for various physical models become exact \cite{Tho}. The analytic treatments of magnetic models \cite{orig},
percolation \cite{Tho,Stauffer_book}, localization \cite{Tho}, and diffusion \cite{RW} on the
Bethe lattice give important physical insights to subsequent developments of the corresponding research fields. 
Therefore, if the critical phenomena of AP on the Bethe lattice is completely understood, one knows MFT for AP exactly. 

One of the theoretical merits of the Bethe lattice is that one can set up exact self-consistent equations on the lattice. 
Recently we have developed an exact self-consistent simulation method for an arbitrary percolation process on
the Bethe lattice \cite{Chae0}. 
From the self-consistent simulation method, we have shown that the phase transition of the Achlioptas-type 
explosive percolation \cite{Ach} undergoes continuous transition regardless of the details of growth rules.
In this paper we analyze the critical properties of AP on the Bethe lattice
by use of the developed self-consistent simulation. 
In the self-consistent simulation the order parameter $P_\infty$ and
the average size $S$ of finite cluster on  the Bethe lattice are directly measured.
Therefore the exponents $\beta$ and $\gamma$ are also obtained directly
without the finite size scaling, and our work can indeed establish exact MFT of AP. 

In addition Lau {\it{et al}}. suggested the modified AP on the $k$-partite graph,
which we call ${\rm{AP}}_k$ \cite{Z2SYM}.
So ${\rm{AP}}_2$ means the original AP on the bipartite graph.
Based on the simple arguments, the transition of ${\rm{AP}}_k$  is conjectured to belong to another new universality class
when the growth of clusters in ${\rm{AP}}_k$ spontaneously breaks the $Z_k$ symmetry of the $k$-partite graph \cite{Z2SYM}.
However the conjecture have never been confirmed quantitatively, yet. 
Therefore, in this paper we also study the MFT of ${\rm{AP}}_3$ by using the triangular cactus structure,
which is an expanded structure of the Bethe lattice and exactly tripartite \cite{Fisher, Tho}.
By the self-consistent simulation we will also find the mean-field exponents $\beta$ and $\gamma$ for ${\rm{AP}}_3$, or $\beta_3$ and $\gamma_3$.
Finally from the results of ${\rm{AP}}_2$ from the Bethe lattice and ${\rm{AP}}_3$ from the triangular cactus,
the inequalities between $\beta_2$ ($\beta$ for ${\rm{AP}}_2$)  and $\beta_3$ and between $\gamma_2$ ($\gamma$ for ${\rm{AP}}_2$) and $\gamma_3$
will be provided in the mean-field level.
From the obtained inequalities, we will suggest the inequalities for all $\beta_k$'s and $\gamma_k$'s.   

This paper is organized as follows. The ordinary AP or ${\rm{AP}}_2$ on the Bethe lattice 
is studied based on the self-consistent simulation in Sec. II. ${\rm{AP}}_3$ on the triangular cactus is defined and studied Sec. III. Finally we summarize our results in Sec. IV.

\section{${\rm{AP}}_2$ on the Bethe lattice}

\begin{figure}[ht]
\includegraphics[width=5cm]{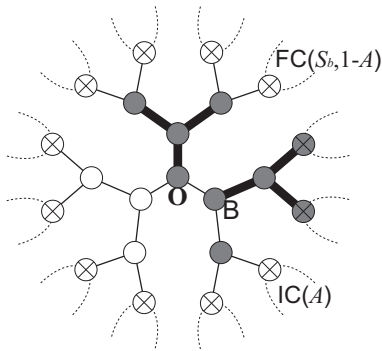}
\caption{Schematic diagram for ${\rm{AP}}_2$ on the the Bethe lattice with $z=3$. The center part consists of a three-generation Cayley tree with edge sites denoted by $\otimes$. Each edge site is connected to an infinite cluster (IC) with the probability $A$ or to a finite cluster (FC) of average size of $S_b$ with the probability $1-A$. Thick lines mean occupied bonds and thin lines mean vacant bonds. If the cluster B is selected, gray sites merge into one cluster.} \label{FIG1}
\end{figure}

In the ordinary AP or ${\rm{AP}}_2$ one cluster is randomly selected instead of a bond or a site 
and the cluster merges all the nearest neighboring clusters to form a new cluster.
This means that in each growth process multiple bonds can be occupied at the same time. Therefore the natural control parameter in ${\rm{AP}}_2$ is the number of clusters per site $n$ instead of the fraction $p$ of the occupied bonds or sites \cite{2Ds,Z2SYM}.

The Bethe lattice is the infinite Cayley tree in which tree structures connected to the center site
{\bf{O}} are identical to one another as schematically shown in \FIG{FIG1} \cite{Tho, Stauffer_book}. The  Bethe lattice is of course bipartite.
Therefore ${\rm{AP}}_2$ on the Bethe lattice is expected to belong to a new universality class different from that of the random percolation.

Let us now briefly explain the self-consistent simulation method for arbitrary percolation on the Bethe lattice with coordination number $z$
in Ref. \cite{Chae0}. In the method we originally used the fraction $p$ of occupied bonds or sites, but we use
the number $n$ of clusters per site in this paper.
To set up self-consistent equations on which the simulation method is based, first consider a part
of the Bethe lattice with $m$ generations from the center site {\bf{O}}, which has  total $N_0 = 1 + z(k^m - 1)/(k - 1)$ sites, where $k = z - 1$.
To make a complete Bethe lattice, one should add an infinite branch to each of $zk^{m-1}$ edge sites.
To calculate the order parameter $P_\infty (n)$ of percolation,
which is defined by the probability for {\bf{O}} to belong to an infinite cluster
at a given $n$,
we need to know the probability $A$ with which an occupied edge site connected to an infinite cluster.
Let $P_{m\infty}(n,A)$ be $P_\infty$ which is calculated from a Bethe lattice
with the $m$ generations from {\bf{O}} and $zk^{m-1}$ infinite branches.
Then the self-consistent equation for $P_\infty$ becomes 
\be \label{SE0}
P_{\infty} = P_{m\infty}(n,A) = P_{m^\prime \infty}(n,A)
\ee
for any combination of $\{m,m^\prime\}$.
Let us define $P_{mst}(n,A,S_b)$ as the probability that a cluster including
{\bf{O}} with $s$ sites and $t$ edge sites occurs within the $m$-generation
tree. Then
\be \label{SE01}
P_{m\infty}(n,A) = 1 - \sum_{t}(1-A)^t \sum_{s} P_{mst}(n,A,S_b)
\ee
where $S_b$ is the average size of the finite cluster connected to an edge site of the $m$-generation tree as shown in Fig. 1. The self-consistent
equation for the average size $S$ of the finite clusters including {\bf{O}} can also be written as
\be \label{SE1}
S = S_m(n,A,S_b) = S_{m^\prime}(n,A,S_b)
\ee
where
\be \label{SE11}
S_m(n,A,S_b) = \frac{\sum_{s,t} P_{mst}(n,A,S_b)[s + tS_b](1 - A)^t}{1 - P_\infty} .
\ee
If one cannot calculate $P_{mst}(n,A,S_b)$ analytically to solve the self-consistent equations,
one should estimate $P_{mst}(n,A,S_b)$   indirectly. One of  such indirect
methods is a simulation method. We have developed  a simulation method
to solve self-consistent equations, which we call the
self-consistent simulation \cite{Chae0}. In the self-consistent simulation, $P_{mst}(n,A,S_b)$ is estimated by the relation $P_{mst}(n,A,S_b) =
N_{mst}(n,A,S_b)/N_{cluster}$, where $N_{mst}(n,A,S_b)$ is the number of clusters including {\bf{O}}
with $s$ sites and $t$ edge sites within the $m$-generation tree that occurred in simulations.
Of course, $N_{cluster}$ is the total number of clusters which includes {\bf{O}} within
the $m$-generation tree that occurred in the same simulation runs. In the simulation both $P_{mst}(n,A,S_b)$ and $P_{m^\prime st}(n,A,S_b)$
are estimated simultaneously using the Bethe lattice with the
$m$-generation tree if $m > m^\prime$.

Since we don't know $A$ and $S_b$ a priori,
the iteration processes are needed in self-consistent simulation.
From initially guessed values for $A$ and $S_b$,
the final or saturated values of $A$ and $S_b$
are obtained by the iteration of unit simulation process.
The unit simulation process consists of the following two steps.
(I) By use of the simulation runs based on the given values $A$ and $S_b$,
$P_{mst}(n,A,S_b)$ and $P_{m^\prime st}(n,A,S_b)$ are
estimated. (II) From the estimated $P_{mst}(n,A,S_b)$, new $A$ and $S_b$
are calculated by utilizing self-consistent equations (\ref{SE0}) and (\ref{SE1}).
In the unit simulation process to get new $A$ and $S_b$, the quantities like $P_{mst}(n,A,S_b)$ are estimated by
averaging over at least $10^6$ simulation runs. 
Such unit process is repeated until $A$ and $S_b$ reach the saturation values.
Using the saturated values of $A$ and $S_b$, $P_{\infty}$ and $S$ are estimated from 
Eqs. (\ref{SE01}) and (\ref{SE11}).
In the self-consistent simulation,
it should be careful to choose $m^\prime(<m)$ for a given $m$ as addressed in Ref. \cite{Chae0}.
If $m^\prime$ is too small, the clusters within the $m^\prime$-generation tree cannot have physical properties of ${\rm{AP}}_2$ enough to give physically plausible solutions for self-consistent equations (\ref{SE0}) and (\ref{SE1}). If $m^\prime$ is very close to $m$, $P_{m\infty}(p,A)$ is numerically not so much distinct from $P_{m^{\prime}\infty}(p,A)$ and the self-consistent equation (\ref{SE0})
hardly gives the physically right solution. 
From the simulations with various sets of $\{m,m^\prime\}$ it is confirmed that suitable
choice of $m^\prime$ should be in the interval $m/3 <m^\prime<m/2$.

\begin{figure}[ht]
\includegraphics[width=8.5cm]{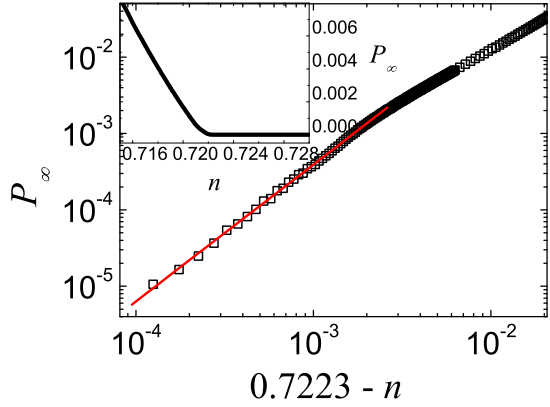}
\caption{
(Color online) Plot of $P_\infty$ for ${\rm{AP}}_2$ against $n_c - n$ with $n_c = 0.7223$. 
The line denotes the relation (\ref{OR}) with $n_c = 0.7223$ and $\beta_2 = 1.79$.
Inset is the raw-data plot of $P_\infty$ against $n$.
} \label{FIG2}
\end{figure}
\begin{figure}[ht]
\includegraphics[width=8.5cm]{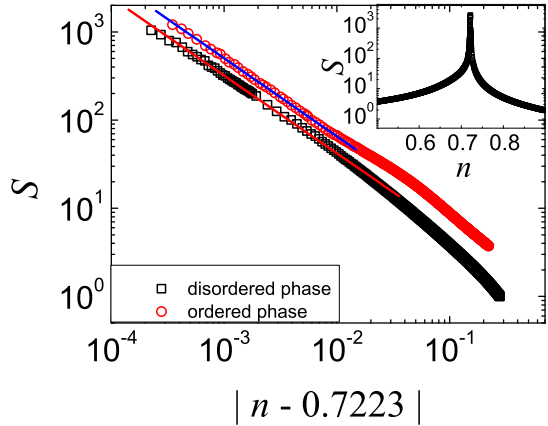}
\caption{(Color online) Plot of $S$ for ${\rm{AP}}_2$ against $|n_c - n|$ for ordered phase ($n<n_c$) and disordered phase ($n>n_c$). The lines denote the relations (\ref{GPMS}) with 
$\gamma_2^- =\gamma_2^+ = 0.88$ and $n_c = 0.7223$. 
Inset is the raw-data plot of $S$ against $n$.
}\label{FIG3}
\end{figure}

The results of the self-consistent simulation with $z=3$, $m=20$, $m^\prime=9$ for ${\rm{AP}}_2$ are displayed in \FIGS{FIG2} and \ref{FIG3}. From the data for $P_\infty$ in \FIG{FIG2} the critical density $n_c$ and the order parameter exponent $\beta_2$ are obtained based on the equation 
\be \label{OR}
P_\infty \simeq (n_c -n )^{\beta_2},
\ee
which holds for the ordered phase 
or for $n<n_c$ near the critical point, i.e., $n \rightarrow n_c^-$. The obtained $n_c$ and $\beta_2$
are $n_c = 0.7223(1)$ and $\beta_2 = 1.79(3)$.
We have checked the results for the simulation for some other combinations ($m=14, m^\prime$) with $m/3 <m^\prime<m/2$ and found the same results.  
For another consistent checks we also applied the self-consistent simulation on the Bethe lattice with $z=6$ to obtain $\beta_2 = 1.79(3)$. These numerical results for $\beta_2$ is close to the previous estimate $\beta_2 = 1.78(8)$ on the ER graph \cite{ERN}, but our estimate has much smaller errors.    

From the data for the average size $S$ of finite clusters in \FIG{FIG3} and the equation
\be \label{GPMS}
S \simeq \left\{ \begin{array}{ll} |n-n_c|^{-\gamma_2^-} & \textrm{if $n_c < n$}\\ |n-n_c|^{-\gamma_2^+} & \textrm{if $n_c > n$} \end {array} \right., 
\ee 
we also estimated $n_c$, $\gamma_2^-$ and $\gamma_2^+$. Obtained $n_c$ is nearly the same as that obtained from the data in 
\FIG{FIG2}. 
We also obtain $\gamma_2=\gamma_2^-=\gamma_2^+ = 0.88(1)$, in which no asymmetry is found between
the disordered phase ($n_c < n$) and the ordered phase $n_c > n$. The result $\gamma_2=0.88(1)$ is also consistent with the previous estimate $\gamma_2=0.88(10)$ on the ER Graph. We also obtained $\gamma_2=0.88(3)$ on the Bethe lattice with $z=6$.
These results clearly show that the obtained values of $\beta_2$ and $\gamma_2$ for $\rm{AP}_2$
are significantly different from those for the random percolation \cite{Stauffer_book}.

In conclusion our estimates $\beta_2=1.79(3)$ and $\gamma_2=0.88(1)$ on the Bethe lattice are far more precise MFT exponents for ${\rm{AP}}_2$ on the bipartite graph, since the dimensionality of the Bethe lattice is infinite. 

\section{${\rm{AP}}_3$ on the triangular cactus}

\begin{figure}[ht]
\includegraphics[width=8.5cm]{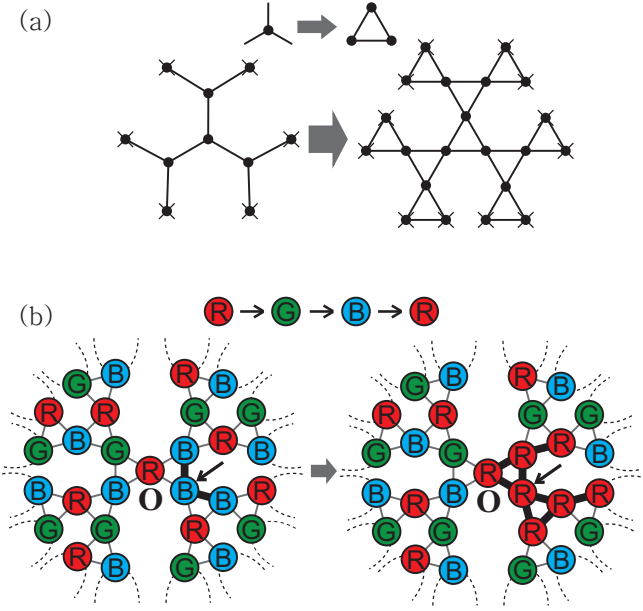}
\caption{(Color online)
(a) Formation of the triangular cactus from the Bethe lattice with $z=3$. In the cactus each site of the Bethe lattice is replaced with a triangle composed of three sites. Each edge site which is denoted by ``X" is connected to an infinite branch.
(b) ${\rm{AP}}_3$ on the triangular cactus. The triangular cactus is tripartite as shown in the figure. If a blue-colored cluster 
indicated by an arrow is selected as in the left figure, it agglomerates all ``R" neighbors and becomes a ``R"-colored cluster by the rule, $\rm{R} \rightarrow \rm{G} \rightarrow \rm{B} \rightarrow \rm{R}$ as in the right figure.} \label{FIG4}
\end{figure}
The triangular cactus was first introduced by Fisher and Essam \cite{Fisher} to investigate the effects of loops \cite{Tho} on the percolation.
As shown in \FIG{FIG4} (a), the triangular cactus with coordination number $z=4$ can be constructed from the Bethe lattice with $z=3$. Each site in the Bethe lattice is replaced with a triangle of three sites to form the triangular cactus as shown in \FIG{FIG4} (a).
Thus the dimensionality of the triangular cactus is infinite as the Bethe lattice.
Moreover, the triangular cactus is exactly tripartite, not bipartite as shown in \FIG{FIG4} (b). Therefore one can expect that the critical phenomena of ${\rm{AP}}_2$ on the triangular cactus belong to the random percolation universality class.

Recently, Lau {\it{et al}}. suggested the modified AP on the $k$-partite graph, which we called ${\rm{AP}}_k$  \cite{Z2SYM}.
It is conjectured that the universality class of ${\rm{AP}}_k$ depends on $k$ \cite{Z2SYM}.
However ${\rm{AP}}_k$ for $k \ge 3$ has never been quantitatively studied, yet.   
In this section ${\rm{AP}}_3$ on the triangular cactus is studied 
to obtain MFT of ${\rm{AP}}_3$.

In ${\rm{AP}}_3$ on the triangular cactus one cluster is randomly selected and the cluster merges some of the nearest
neighboring clusters into a new cluster, instead of all neighboring clusters in ${\rm{AP}}_2$ on bipartite graph.
In a tripartite graph, initially, three colors are arranged such that no pair of nearest neighbor sites has the same color.
Therefore we can identify the cluster by colors such as red (R), green (G), and blue (B).
${\rm{AP}}_3$ is defined such that a selected cluster with ``R" can join only with neighbors of the color ``G",
``G" can join only with neighbors of the color ``B", ``B" can join only with neighbors ``R",
based on a cyclic rule, $\rm{R} \rightarrow \rm{G} \rightarrow \rm{B} \rightarrow \rm{R}$ \cite{Z2SYM}.
For example, a certain cluster with ``R" is selected, then the cluster merges 
all neighboring clusters colored by ``G" into a new cluster, and the merged cluster becomes
a new ``G"-colored cluster from the rule, $\rm{R} \rightarrow \rm{G} \rightarrow \rm{B} \rightarrow \rm{R}$. One can apply the $\rm{R} \rightarrow \rm{B} \rightarrow \rm{G} \rightarrow \rm{R}$ rule to the model, but it cannot be physically different from the model with the $\rm{R} \rightarrow \rm{G} \rightarrow \rm{B} \rightarrow \rm{R}$ rule.

For  MFT of ${\rm{AP}}_3$ on the tripartite graph, we use the self-consistent simulation method for ${\rm{AP}}_3$ on the triangular cactus. The self-consistent simulation is almost the same as that for ${\rm{AP}}_2$ on the Bethe lattice.
First consider a part of the triangular cactus with $m$-generations from {\bf{O}},
which has total number of sites $N_0 = 2^{m+2}-3$.
To make a complete triangular cactus, one should add an infinite branch to each of $2^{m+1}$ edge sites.
Other details of the self-consistent simulation on the triangular cactus are exactly the same as those on the Bethe lattice.
For instance, Eqs. (\ref{SE0})-(\ref{SE11}) are the self-consistent equations not only for the Bethe lattice but for the triangular cactus.

We first checked the results of self-consistent simulation with $m=20, m^\prime = 9$ for AP 
on the triangular cactus.
From the data for $P_\infty$ and $S$, $n_c$, $\beta$ and $\gamma$ is estimated as $n_c = 0.6761(1)$, $\beta = 1.01(2)$, and $\gamma = 1.00(2)$. 
Since the triangular cactus is not bipartite but tripartite, this result supports that the critical phenomena of ordinary AP 
on the tripartite graph belong to the random percolation universality class with $\beta=\gamma=1$ as expected in Ref. \cite{Z2SYM}.

\begin{figure}[ht]
\includegraphics[width=8.5cm]{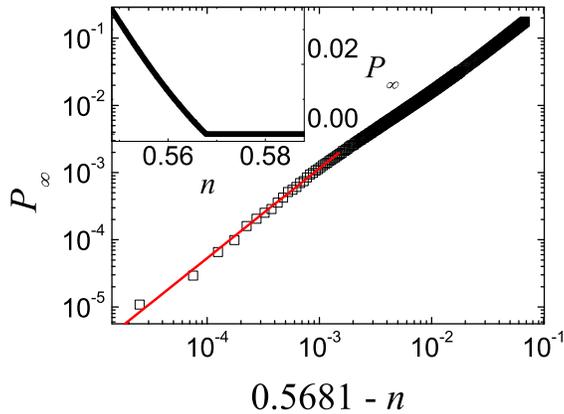}
\caption{(Color on line) Plot of $P_\infty$ for ${\rm{AP}}_3$ on the triangular cactus against $n_c - n$ with $n_c = 0.5681$.
The line denotes the relation similar to (\ref{OR}) with $n_c = 0.7223$ and $\beta_3 = 1.35$. Inset is the raw-data plot of $P_\infty$ against $n$.
} \label{FIG5}
\end{figure}
\begin{figure}[ht]
\includegraphics[width=8.5cm]{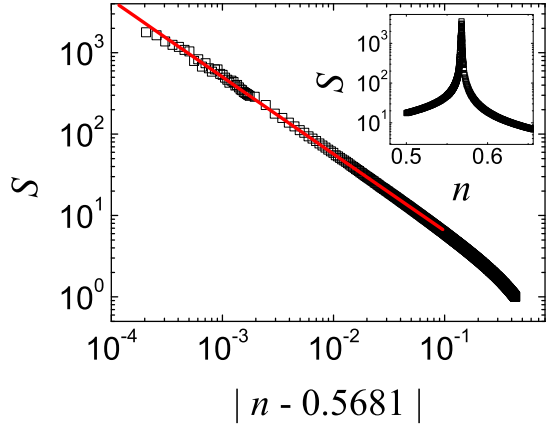}
\caption{(Color on line) Plot of $S$ for ${\rm{AP}}_3$ on the triangular cactus against $n_c - n$ with with $n_c = 0.5681$ for disordered phase ($n> n_c$).
$S$ for ordered phase ($n< n_c$) is not shown, because it behaves almost the same as that for disordered phase. The line denotes the relation similar to Eq. (\ref{GPMS}) with $\gamma_3^- = 0.94(2)$. Inset is the raw-data plot of $S$ against $n$.} \label{FIG6}
\end{figure}

On the other hand, the critical phenomena of ${\rm{AP}}_3$ is different from AP on the triangular cactus.
The results of the same self-consistent simulation for ${\rm{AP}}_3$ on the triangular cactus are
shown in \FIGS{FIG5} and \ref{FIG6}.
By using the similar equations to Eqs. (\ref{OR})
and (\ref{GPMS}) we have obtained the order parameter
exponent $\beta_3$ and the susceptibility exponent 
$\gamma_3$ for ${\rm{AP}}_3$ on the triangular cactus. The results are $\beta_3 = 1.35(5)$ and $\gamma_3 = 0.94(2)$ with
$n_c = 0.5681(1)$ as shown in \FIGS{FIG5} and \ref{FIG6}. In conclusion our estimated $\beta_3$ and $\gamma_3$ are the first MFT exponents for ${\rm{AP}}_3$, since the dimensionality of the triangular cactus is infinite. 
 
The obtained exponents $\beta_3$ and $\gamma_3$
satisfy the relations $\beta_\infty< \beta_3 < \beta_2$ and $\gamma_\infty > \gamma_3 > \gamma_2$, where $\beta_\infty(=1)$ and $\gamma_\infty(=1)$ are the MFT exponents of the random percolation.
The relations of exponents suggests that the MFT exponents of ${\rm{AP}}_k$ on the $k$-partite graph  approach to those of the random percolation as $k$ increases.

\section{Summary}
Finding the exact MFT is the first step to understand the various physical properties of a new model.
AP was suggested as a natural extension of the standard percolation model.
Some numerical studied for AP have been done on lower-dimensional lattices and random graphs \cite{1DRING,CRIT,ERN,Z2SYM}.
Based on those numerical studies, it was conjectured that  AP belongs to a
new universality class if the growth of cluster breaks $Z_k$ symmetry on $k$-partited graph.
However, the mean-field approach based on the evolutionary dynamics of clusters did not agree with the numerical simulations.
This strongly indicates that AP is not fully understood even in mean-field level \cite{ERN}.
Therefore, in order to provide an exact MFT, we apply the self-consistent simulation method \cite{Chae0}
to APs on the Bethe lattice and triangular cactus.
From the direct and precise measurement of $P_\infty$ and $S$ through the self-consistent simulation,
we obtain $\beta_2=1.79(3)$ and $\gamma_2=0.88(1)$ on the Bethe lattice when $Z_2$ symmetry is broken
spontaneously at the transition threshold.
Similarly, we obtain $\beta_3=1.35(5)$ and $\gamma_3=0.94(2)$ on triangular cactus if $Z_3$ symmetry is broken spontaneously.
However, since the triangular cactus is not bipartite, $\rm{AP}$ model on triangular cactus gives
$\beta=1.01(2)$ and $\gamma=1.00(2)$. This result shows that ordinary $\rm{AP}$
on triangular cactus belongs to the same universality class with ordinary percolation.
Therefore, the results for ${\rm{AP}}_3$ on triangular cactus provide the exact MFT verifying the Lau {\it{et al.}}'s arguments \cite{Z2SYM}.
In addition, by comparing the obtained critical exponents with those of random percolation, we
also find the inequalities $\beta_\infty <\beta_3<\beta_2$ and $\gamma_{\infty}>\gamma_3>\gamma_2$. These inequalities
also quantitatively verify the conjecture that the universality class of ${\rm{AP}}_k$ depends on $k$ \cite{Z2SYM}.

\begin{acknowledgements}
This work was supported by National Research Foundation
of Korea (NRF) Grant funded by the Korean Government
(MEST) (Grants No. 2011-0015257)
and by Basic Science Research Program 
through the National Research Foundation of Korea(NRF) funded by
the Ministry of Education, Science and Technology (No. 2012R1A1A2007430).
\end{acknowledgements}

\end{document}